\documentclass{aa}

\usepackage{graphicx}
\usepackage{multirow}
\usepackage{tabularx}
\usepackage{longtable,tabu}
\usepackage{lscape}
\usepackage{subfigure}
\usepackage{natbib}
\usepackage[dvipsnames]{xcolor}
\bibpunct{(}{)}{;}{a}{}{,} 
\def\logR{\ensuremath{\log R^{\prime}_{\mathrm{HK}}}}

\begin{document}

\title{Ca\,{\sc ii}\,H\&K stellar activity parameter: a proxy for stellar Extreme Ultraviolet Fluxes}

\author{A. G. Sreejith\inst{1} \and L. Fossati\inst{1} \and A. Youngblood\inst{2} \and K. France\inst{2} \and S. Ambily\inst{2}}
\institute{Space Research Institute, Austrian Academy of Sciences, Schmiedlstrasse 6, 8042 Graz, Austria\\
\email{sreejith.aickara@oeaw.ac.at}
            \and
            Laboratory for Atmospheric and Space Physics, University of Colorado, UCB 600, Boulder, CO, 80309, USA}

\date{Received date /
Accepted date }

\abstract
{Atmospheric escape is an important factor shaping the exoplanet population and hence drives our understanding of planet formation. Atmospheric escape from giant planets is driven primarily by the stellar X-ray and extreme-ultraviolet (EUV) radiation. Furthermore, EUV and longer wavelength UV radiation power disequilibrium chemistry in the middle and upper atmosphere. Our understanding of atmospheric escape and chemistry, therefore, depends on our knowledge of the stellar UV fluxes. While the far-ultraviolet fluxes can be observed for some stars, most of the EUV range is unobservable due to the lack of a space telescope with EUV capabilities and, for the more distant stars, to interstellar medium absorption. Thus, it becomes essential to have indirect means for inferring EUV fluxes from features observable at other wavelengths. We present here analytic functions for predicting the EUV emission of F-, G-, K-, and M-type stars from the log\,$R'_{HK}$ activity parameter that is commonly obtained from ground-based optical observations of the Ca\,{\sc ii}\,H\&K lines. The scaling relations are based on a collection of about 100 nearby stars with published log\,$R'_{HK}$ and EUV flux values, where the latter are either direct measurements or inferences from high-quality far-ultraviolet (FUV) spectra. The scaling relations presented here return EUV flux values with an accuracy of about three, which is slightly lower than that of other similar methods based on FUV or X-ray measurements.}


\maketitle
\section{Introduction}\label{sec:intro}
%
Planet atmospheric escape is one of the most important processes affecting the evolution of planetary atmospheres \citep[e.g.,][]{lopez2013,owen2013,jin2014} and it has played a key role in shaping the atmospheres of the inner solar system planets \citep[e.g.,][]{lammer2018,lammer2020}. It is believed that escape also has a profound impact on the observed exoplanet population \citep[e.g.,][]{owen2017,jin2018}. Furthermore, because of the difficulty of directly observing and studying young planets, constraining atmospheric accretion processes requires tracing the evolution, and hence mass loss, of older planets, which can instead be more easily observationally characterised \citep[e.g.,][]{jin2014,kubyshkina2018,Kubyshkina2019a,Kubyshkina2019b}. 

The vast majority of the known exoplanets orbit close to their host stars and are therefore strongly irradiated. Except in specific cases \citep[e.g., very young or atmosphereless planets;][]{owen2016,fossati2017,vidotto2018}, escape is mainly driven by heating due to absorption of the stellar high-energy radiation, in particular extreme ultraviolet (EUV) and X-ray photons, altogether called XUV \citep[e.g.,][]{watson1981,lammer2003,yelle2004,murray2009,koskinen2013a,koskinen2013b}. XUV photons can heat the thermosphere to temperatures of the order of 10$^4$\,K, which causes the atmosphere to expand, possibly hydrodynamically, leading to mass loss. For classical hot Jupiters (e.g., HD209458b), mass-loss rates are of the order of 10$^{9-10}$\,g\,s$^{-1}$, but they can become significantly larger for planets orbiting hot stars \citep[e.g.,][]{fossati2018b,garcia2019,hiroto2020}, or for planets orbiting young, active stars \citep[e.g.,][]{penz2008,kubyshkina2018a}, or for low-gravity planets \citep[e.g.,][]{lammer2016,cubillos2017}.

Atmospheric escape has been directly observed for a few close-in exoplanets \citep[e.g.,][]{vidal2003,fossati2010,linsky2010,lecavelier2012,haswell2012,ehrenreich2015,bourrier2018,mansfield2018,sing2019,cubillos2020} and predicted for many others \citep[e.g.,][]{ehrenreich2011,salz2016}, however, to extract the relevant information from the observations and/or to theoretically estimate the physical conditions of planetary upper atmospheres and infer mass-loss rates, it is necessary to quantify the amount of XUV flux irradiating the planet. Because of the lack of an observational facility with adequate capabilities \citep{france2019} and interstellar medium absorption for farther stars, it is not possible to directly observe the EUV stellar emission for stars other than the Sun. Although there are a few space telescopes with X-ray capabilities, the X-ray stellar emission is typically an order of magnitude less intense than the EUV emission and it has a smaller absorption cross-section in a hydrogen-dominated atmosphere \citep[e.g.,][]{cecchi2009,Sanz-Forcada2011,tu2015}, making the X-ray fluxes alone inadequate to constrain upper atmospheres and escape.

Several methods and scaling relations have been devised to estimate stellar XUV fluxes from proxies observable at longer/shorter wavelengths, such as X-ray and ultraviolet (UV) fluxes \citep[e.g.,][]{Sanz-Forcada2011,linsky2014,Chadney2015,louden2017,france2018}. For example, \citet{linsky2014} derived a correlation between EUV and Ly$\alpha$ emission fluxes to use observations of the Ly$\alpha$ line to infer stellar XUV fluxes in different bands. In their work, \citet{linsky2014} employed a mixture of EUV spectra measured in the past for a few nearby stars by the EUVE satellite and solar spectral synthesis computations. \citet{linsky2013} further derived similar correlations between the fluxes of the Ly$\alpha$ line and several emission features in the ultraviolet (C{\sc ii}, C{\sc iv}, O{\sc i}, Mg{\sc ii}\,h\&k lines) and optical (Ca\,{\sc ii}\,H\&K lines) bands. However, these scaling relations require either reconstructing the Ly$\alpha$ line, whose core is typically heavily absorbed by the interstellar medium, employing two scaling relations to derive the stellar XUV fluxes from emission lines other than Ly$\alpha$. More recently, \citet{france2018} followed the same strategy of \citet{linsky2014} by deriving a correlation between EUV emission fluxes in two bands (i.e., 90--360\,\AA\ and 90--911\,\AA) and lines in the far-ultraviolet (N{\sc v} and Si{\sc iv}) for which high-quality spectra had been collected with the Hubble Space Telescope. The advantage of the relation of \citet{france2018} over that of \citet{linsky2013,linsky2014} is the larger sample and the use of lines forming at temperatures closer to that of the EUV formation temperature range. Other works have reconstructed stellar EUV spectra by scaling the solar UV spectrum to match measurements of high-energy far-ultraviolet (FUV) emission lines \citep[e.g.,][]{fossati2015,fossati2018}. FUV and X-ray measurements have also been combined via the differential emission measure \citep[e.g.,][]{louden2017} or by scaling the observed XUV fluxes of nearby stars \citep[e.g.,][]{Chadney2015}.

All of these EUV estimation methods require space-based observations at FUV and/or X-ray wavelengths, which are not straightforward to obtain, particularly for large samples of stars. One way around this problem is to employ the stellar rotation rate as a proxy for activity, hence XUV emission, of late-type stars \citep[e.g.,][]{johnstone2015,tu2015}, but this method is less accurate than those based on the direct measurement of chromospheric and/or coronal emission and it has never been empirically verified, except for Sun-like stars \citep[e.g.,][]{ribas2005}. Furthermore, in some cases, this method may lead to wrong answers, such as for the planet-hosting star WASP-18, which is a fast rotator, but has an extremely low activity level that is believed to be dampened by tidal interactions with the massive close-in planet \citep{fossati2018}.

We employ here the stellar sample of \citet{france2018} to derive the correlation between the \logR\ stellar activity index and EUV fluxes in the 90--911\,\AA\ wavelength range. The \logR\ index is a measure of the stellar chromospheric emission flux at the core of the deep Ca\,{\sc ii}\,H\&K absorption lines ($\sim$3933 and $\sim$3968\,\AA), which has the advantage of lying in the optical band and therefore of being easily observable from the ground. However, \logR\ has the disadvantage of forming mostly in the chromosphere, hence at lower temperatures compared to the typical formation temperature of the EUV stellar emission. A large number of studies have and continue to make use of the \logR\ index, which has its roots in the Mount Wilson $S$-index ($S_{\rm MW}$), to study stellar activity \citep[e.g.,][]{wilson1978,noyes1984,Duncan1991,Baliunas1995}. In this work, we obtain scaling relations enabling one to infer EUV fluxes for M-, K-, G-, and F-type stars directly from the \logR\ index. This allows one to infer in a simple, direct way the EUV emission of a large number of late-type stars without the need for high-quality space-based observations.

This paper is organised as follows. Section~\ref{sec:target} presents the considered sample of stars employed to derive the \logR$-$EUV correlation. 
Section~\ref{sec:results} presents the \logR$-$EUV correlation, while we discuss the results and gather the conclusions of this work in Section~\ref{sec:discussion}.

\section{Stellar sample and stellar parameters} \label{sec:target}
We started from considering the stars comprising the sample of \citet{france2018}, 104 F-, G-, K-, and M-type stars lying within about 50\,pc. For each star, \citet{france2018} estimated the EUV emission flux in the 90-360\,\AA\ wavelength range on the basis of FUV observations, in particular of the N{\sc v} and Si{\sc iv} lines, and the stellar bolometric flux at Earth (F$_{\rm bol}$). These EUV estimates have accuracy about a factor of 2 and are based on relationships between the fractional FUV emission line luminosity and the fractional EUV luminosity they obtained from moderate-to-high quality EUV spectra in the 90--360\AA\ wavelength range collected by the EUVE satellite. 

\begin{figure}[h]
\begin{center}
\includegraphics[width=\hsize]{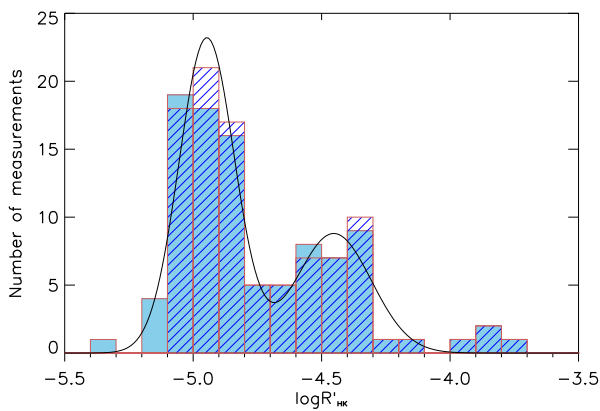}
\caption{Distribution of the median \logR\ values for all stars considered in this work before (blue shaded; 99 stars, 5 stars from our initial sample had no \logR\ values in the literature) and after (hatched; 96 stars) outliers removal, namely following removal of the values lying below the basal level and following the sigma clipping based on the MAD. The black solid line shows the double Gaussian fit to the \logR\ distribution obtained following outliers removal.}
\label{fig:fig1}
\end{center}
\end{figure}

For each star in the sample of \citet{france2018}, we collected from the literature measurements of the \logR\ activity parameter and retained only the stars for which we found at least one measurement \citep{issacson2010,borosaikia2018,jenkins2011,Jenkins2006,Caillault1991,Mamajek2008,gray2006,gray2003,Moutou2014,Arriagada2011,Robertson2012,Anderson2014,Wittenmyer2009,Laws2003,henry1996,CantoMartins2011,pace2013,Astudillo2017,hinkel2017,ment2018,wright2004,Hall2009,Morris2017}. The total number of collected \logR\ measurements is 498. The distribution of the median \logR\ values used in this work is shown in Figure~\ref{fig:fig1}. The distribution appears to be bimodal, with peaks at \logR\ values of about $-$5.0 and $-$4.5, and hence it is similar to the overall distribution of \logR\ values of stars in the solar neighbourhood \citep[e.g.,][]{vaughan1980,gray2003}.


For each star, we identified and removed outlier \logR\ values by employing a two-step approach. First, we removed all \logR\ values lower than $-5.1$ (total of 28 removed measurements for 17 different stars of spectral type G, K, and M), which is the minimum \logR\ value (i.e., usually called ``basal level'') possible for main sequence late-type stars \citep{wright2004}. However, we remark that the sample of \citet{wright2004} was composed mostly of F-, G-, and K-type stars and that several M dwarfs present \logR\ values below the basal level of $-$5.1 \citep[e.g.,][]{Astudillo2017}, which may be due to the fact that \citet{Astudillo2017} focused on planet-hosting stars that are typically inactive because of selection biases and/or that M-type stars may have a basal level different from that of hotter solar-like stars. Then, we excluded further outliers by identifying the \logR\ values deviating more than three times the median absolute deviation (MAD) from the median value. We then calculated the average and median from the remaining \logR\ measurements (Table~\ref{table2}). Figure~\ref{fig:fig1} presents the distribution of median \logR\ values obtained for each star following the removal of the outliers and compares it with the original distribution. It shows that the removal of the outliers has not significantly modified the underlying distribution, which remains bimodal. We confirmed the bimodality of the distribution by fitting it employing a double-peaked Gaussian (shown in Fig.~\ref{fig:fig1}) and by running a Kolmogorov-Smirnov test on the two resultant Gaussian distributions obtaining at high significance that they are indeed drawn from distinct samples. Furthermore, because of the close distance of the stars in our sample, the \logR\ values are not systematically depressed by interstellar medium absorption \citep{fossati2017b}. This process led to a sample comprising 96 stars.

\begin{figure}[h]
\begin{center}
\includegraphics[width=\hsize]{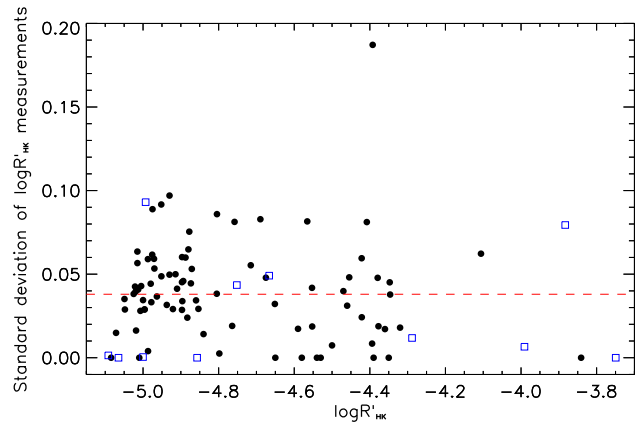}
\caption{Standard deviation obtained from the \logR\ values collected from the literature following outlier removal as a function of the median \logR\ value. Black dots are for F-, G-, and K-type stars, while blue open squares are for M-type stars. The dashed red line indicates the median standard deviation that we adopt as uncertainty on the \logR\ value for stars for which only one \logR\ measurement is available.}
\label{fig:fig2}
\end{center}
\end{figure}

Figure~\ref{fig:fig2} presents the standard deviation of the \logR\ values for each star, following the removal of the outliers, as a function of the median \logR\ value. Except for one star, HD 106516, the standard deviation is smaller than 0.1 and the median value of the standard deviation is about 0.04. Since the scatter on the standard deviation does not increase with increasing \logR\ value, the scatter is most likely driven by measurement uncertainties, rather than by stellar variability. Therefore, for each star, we took the standard deviation as the uncertainty on the median \logR\ value and, for the stars with only one measured \logR\ value, we considered the median value of the standard deviation (i.e., $\sim$0.04) as the uncertainty on the measured \logR\ value. By taking the median \logR\ value, we mitigate the effects on the results of intrinsic stellar variability and non-simultaneity of the \logR\ and EUV measurements.

\begin{figure}[h]
\begin{center}
\includegraphics[width=\hsize]{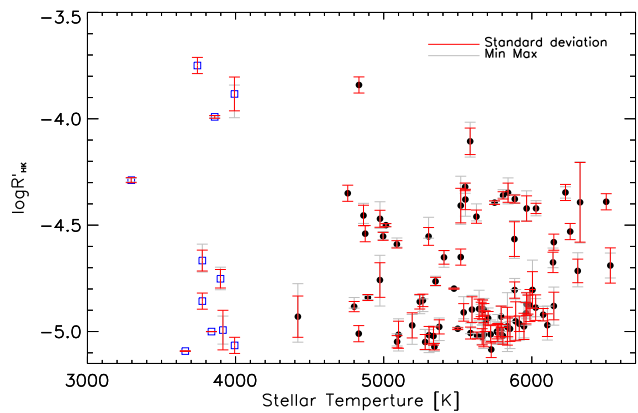}
\caption{Median \logR\ values following outliers removal as a function of stellar effective temperature. Black dots are for F-, G-, K-type stars and blue open squares are for M-type stars. The red error bars indicate the standard deviation of the \logR\ measurements, while the grey error bars indicate the minimum and maximum \logR\ values collected from the literature and following outliers removal. For stars with a single \logR\ measurement, the standard deviation is the median value of the standard deviation obtained for all stars in the sample and shown by the red dashed line in Figure~\ref{fig:fig2}.}
\label{fig:fig3}
\end{center}
\end{figure}

Figure~\ref{fig:fig3} presents the distribution of the median \logR\ values as a function of stellar effective temperature (T$_{\rm eff}$) obtained from Gaia DR2 catalogue \citep{gaiadr2}. Stellar temperature values from \citet{france2018} were retained for stars where temperature information was absent in the Gaia DR2 catalogue. This plot further shows the uncertainty and the minimum and maximum values associated with each point. As expected, following the outlier removal based on the MAD, for most stars the median \logR\ value lies roughly in the middle between the minimum and maximum values. The majority of the stars are G-type, but also M- and late F-type are well represented, though early and mid K-type stars are sparsely represented.

For each considered star, we updated T$_{\rm eff}$, the stellar distance (d), and stellar radius, hence the bolometric flux at Earth, using the values given in the Gaia DR2 catalogue \citep{gaiadr2}. The stellar radii were updated using the Gaia stellar magnitudes, effective temperatures, and distances as follows
\begin{equation}\label{eq:radius}
    R_{\rm star}\,=\,R_{\rm sun}\,\frac{T_{\rm eff,sun}^2}{T_{\rm eff,star}^2}\,\frac{d_{\rm star}}{d_{\rm sun}}\,10^{\frac{m_{\rm sun}-m_{\rm star}}{5}}\,,
\end{equation}
where $R_{\rm star}$ is the stellar radius, $R_{\rm sun}$ is the solar radius, $T_{\rm eff,sun}$ is the solar effective temperature (5777\,K), $T_{\rm eff,star}$ is the stellar effective temperature, $d_{\rm star}$ is the distance to the star, $d_{\rm sun}$ is the distance to the Sun, $m_{\rm sun}$ is the solar apparent $V$-band magnitude ($-$26.73), and $m_{\rm star}$ is the apparent stellar $V$-band magnitude. This equation assumes that the stellar apparent magnitude is not affected by interstellar extinction, which we assume to be the case given the close distance to the stars in our sample. 

\begin{figure}[h]
\begin{center}
\includegraphics[width=\hsize]{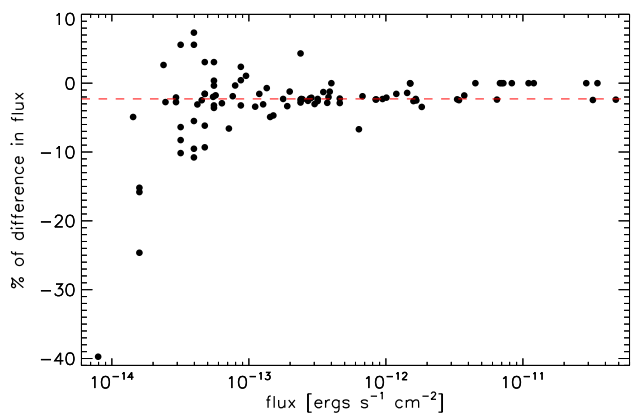}
\caption{The difference in EUV flux at Earth values in the 90--360\,\AA\ wavelength range between those given by \citet{france2018} and those obtained employing the scaling relation in \citet[][Eq.~(4)-(6)]{france2018}} and the stellar parameters derived from the GAIA DR2 catalogue. The median difference is about 2\% (red dashed line). Except for stars with low EUV fluxes (below $2\times10^{-14}$\,erg\,cm$^{-2}$\,s$^{-1}$) the difference is within 10\%.
\label{fig:fig4}
\end{center}
\end{figure}

Our reference sample of stars is that of \citet{france2018}, which is also our source for the stellar EUV fluxes at Earth. However, for most stars, we updated both $T_{\rm eff}$ and radius, hence bolometric fluxes at Earth. For this reason, except for the stars for which the EUV flux originates from EUVE observations, we employed the scaling relations of \citet[][their Equations~(4), (5), and (6)]{france2018} and their N{\sc v} flux measurements to update the stellar EUV fluxes that are listed in Table~\ref{table2}. For the stars without a N{\sc v} flux measurement, we employed the Si{\sc iv} flux measurement. Figure~\ref{fig:fig4} shows the difference between the EUV fluxes given by \citet{france2018} and those we obtained following the update of the bolometric fluxes at Earth. The median difference is about 2\% and, for most stars, the difference lies within 10\%.

\begin{figure*}[t]
\begin{center}
\resizebox{\hsize}{!}
{\includegraphics[width=\textwidth]{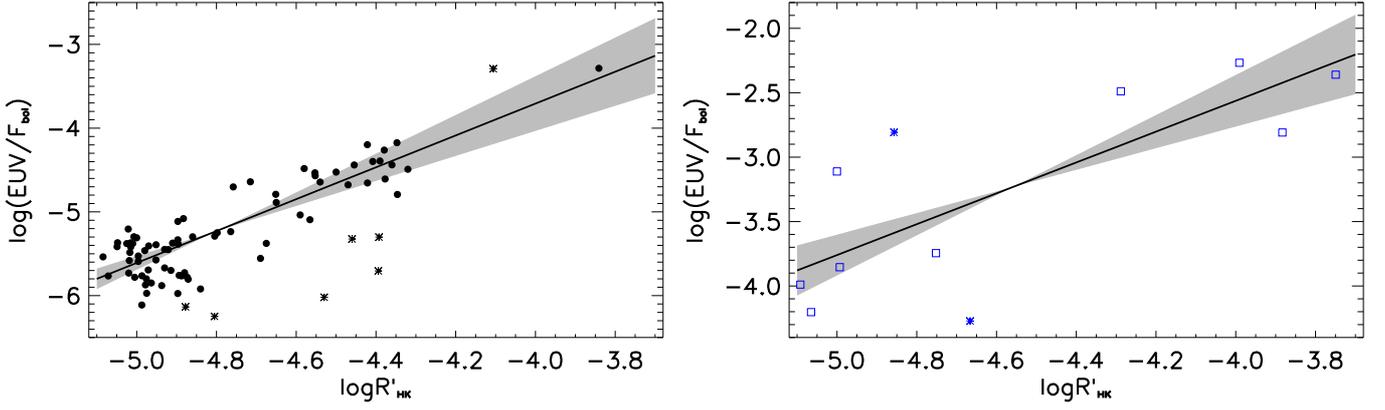}}
\caption{Correlation between the stellar activity index (\logR) and EUV flux for F-, G-, K-type stars (left; black dots) and M-type stars (right; blue open squares). The RMS on $\log (EUV/F_{\rm bol})$ after the fit for F-, G-, K-type stars and for M-type stars is 0.40 and 0.48, respectively. The stars removed as a result of the sigma clipping (see text) are indicated by the asterisks. The grey areas indicate the uncertainties on the fits.}
\label{fig:fig5}
\end{center}
\end{figure*}


\section{Results}\label{sec:results}

\begin{figure*}[h]
\begin{center}
\resizebox{\hsize}{!}
{\includegraphics[width=\textwidth]{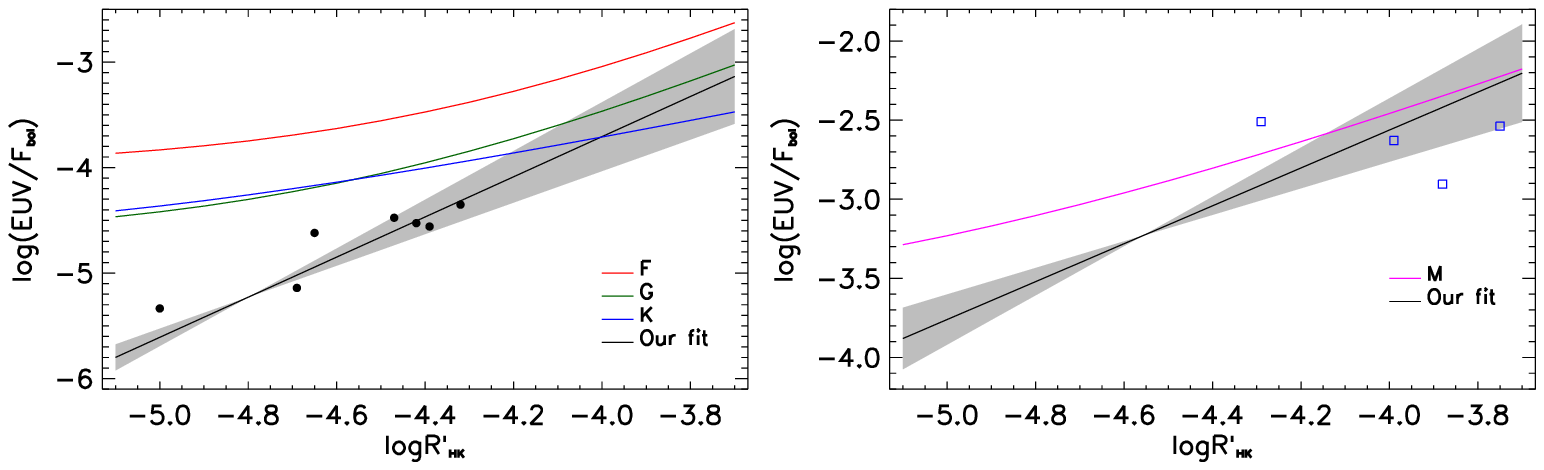}}
\caption{Correlation between EUV and \logR\ activity parameter following the conversion of \logR\ value into Ca\,{\sc ii}\,H\&K chromospheric emission fluxes and the scaling relations of \citet{linsky2013} and \citet{linsky2014} for F- (red), G- (green), K- (blue), and M-type (magenta) stars. The black line and grey shaded area are the linear fits obtained for F-, G-, K-type stars (left) and for M-type stars (right) and shown in Figure~\ref{fig:fig5}. The black dots (left) and blue open squares (right) indicate the position of the stars in common between \citet{linsky2014} and our sample where $\log (EUV/F_{\rm bol})$ are from \citet{linsky2014}. For inactive stars, deriving the EUV fluxes from the \logR\ values employing the scaling relations of \citet{linsky2013} and \citet{linsky2014} leads to overestimations of about one order of magnitude.}
\label{fig:fig6}
\end{center}
\end{figure*}

Figure~\ref{fig:fig5} shows the stellar EUV flux as a function of the median \logR\ value for the stars in our sample and the best linear fit through these points. The linear fit was achieved with a minimized chi-square approach accounting for the uncertainties on both EUV fluxes and \logR\ values. To improve the robustness of the fits, we employed an iterative sigma clipping algorithm, in which we iteratively removed from the fit stars deviating more than 1.5 times the root mean square (RMS) value. We ran separate fits for stars belonging to different spectral types obtaining that the EUV flux vs \logR\ value correlation is comparable for F-, G-, and K-type stars, hence we considered all of them for the joint fit, but the M-type stars appear to follow a different scaling relation. However, we remark that M dwarfs later than M3.5 are fully convective and may not behave like the earlier M dwarfs in terms of chromospheric and coronal emission, but the too small sample does not allow us to identify statistically significant differences when performing separate fits.

We find that the correlation between the stellar activity index (\logR) and the EUV flux in the 90$-$911\,\AA\ wavelength range can be described by a linear fit of the form 
\begin{equation}
\label{eq:our_result}
    \log_{10}{(EUV(90-911\,\AA)/F_{\rm bol})}=c_1 \times \logR\ +c_2\,,
\end{equation}
where the $c_1$ and $c_2$ coefficients are listed in Table~\ref{table:1}. We further compute the Pearson correlation coefficients (PCC) for \logR\ vs $\log (EUV/F_{\rm bol})$, obtaining that the linear correlations are indeed significant (see values in Table~\ref{table:1}). The results indicate that the correlation for F-, G-, and K-type stars is significantly steeper than that for M-type stars, which is in agreement with what has been previously found \citep[see e.g.,][]{linsky2013,linsky2014,france2018}. These results indicate also that our scaling relations provide EUV fluxes with an accuracy of about a factor of three, which is slightly higher than that of other similar methods based on FUV or X-ray measurements. The scatter present in the line fits has the largest contribution to the uncertainties in the EUV flux estimates. 

\begin{table}
\caption{ Parameters of the linear correlation between stellar activity index (\logR) and EUV flux in the 90$-$911\,\AA\ wavelength range described by Eq.~(\ref{eq:our_result}). Column four gives the root mean square (RMS) of the correlation, while column five gives that Pearson correlation coefficients (PCC).}     
\label{table:1}
\centering                                      
\begin{tabular}{c c c c c}          
\hline\hline                        
Sp. Type & c$_1$ & c$_2$ & RMS & PCC \\    
\hline                                  
    F, G, K & $1.90\pm0.41$ & $3.90\pm1.96$  & 0.40 & 0.8748\\      
    M & $1.20\pm0.36$ &  $2.23\pm1.64$  & 0.48 & 0.8753\\
 \hline                                            
\end{tabular}
\end{table}

\section{Discussion and conclusion}\label{sec:discussion}

We presented here a linear correlation between the \logR\ stellar activity index and the EUV flux emitted by late-type stars in the 90$-$911\,\AA\ wavelength range, which is responsible for most of the heating in upper planetary atmospheres. This correlation enables one to convert the chromospheric emission at the core of the Ca\,{\sc ii}\,H\&K lines, parameterised by the \logR\ value and measurable from the ground, into stellar high-energy emission that can then be used to estimate for example exoplanetary atmospheric mass-loss rates.

Before this work, one could have also combined the scaling relations published by \citet{linsky2013} and \citet{linsky2014} to convert the chromospheric emission at the line core into EUV fluxes, though this would have implied first converting the \logR\ value into Ca\,{\sc ii}\,H\&K chromospheric emission and then passing through the estimation of the Ly$\alpha$ fluxes. Therefore we compared the \logR\ vs EUV fluxes obtained following that approach as well as this work's method presented in Section~\ref{sec:results}. To this end, we first converted the \logR\ values into disk-integrated chromospheric emission flux in erg\,cm$^{-2}$\,s$^{-1}$ following \citet{fossati2017b} and \citet{sreejith2019}. In short, the disk-integrated Ca\,{\sc ii}\,H\&K chromospheric line emission at a distance of 1\,AU ( $F_{\rm HK}$(1 AU)) is \citep[see Sect.~2 of][]{fossati2017b}
\begin{equation}\label{eq:E}
F_{HK}(1 AU)=\frac{(S_{\rm MW}\,10^{8.25-1.67\,(B-V)} - 10^{\rm p})\,R_{\rm star}^2}{AU^2}\,,
\end{equation}
where $S_{\rm MW}$ is the $S$-index activity indicator in the Mount-Wilson system \citep{noyes1984,mittag2013}, $R_{\rm star}$ is the stellar radius in cm, and $AU$ is one astronomical unit in cm. The exponent p in Equation~(\ref{eq:E}) is equal to $7.49-2.06\,(B-V)$ for main sequence stars with $0.44\leq B-V < 1.28$ and is equal to $6.19-1.04\,(B-V)$ for main sequence stars with $1.28\leq B-V < 1.60$ \citep{mittag2013}, while the $S_{\rm MW}$ index is defined as
\begin{equation} 
S_{\rm MW} = \frac{ 10^{\log{R^{\prime}_{\rm HK}}} + \frac{10^{\bf{\rm p}}}{\sigma T_{\rm eff}^4} }{1.34 \times 10^{-4}\,C_{cf}}\,,
\end{equation}
where $ C_{cf}$ is \citep{rutten1984}
\begin{equation}
\log_{10}{ C_{cf}} = 0.25\,(B-V)^3 - 1.33\,(B-V)^2 + 0.43\,(B-V) + 0.24\,.
\end{equation}
The parameters employed to derive $F_{\rm HK}$(1 AU) are listed in Table~\ref{table2} and the $B-V$ values have been obtained from the stellar effective temperature by interpolating Table~5\footnote{See also {\tt http://www.pas.rochester.edu/$\sim$emamajek/\\EEM\_dwarf\_UBVIJHK\_colors\_Teff.txt}\,.} of \citet{pecaut2013}.

We then set an array of \logR\ values, converted them to $F_{\rm HK}$(1 AU) as described above, and used the scaling relations of \citet{linsky2013} to obtain the Ly$\alpha$ fluxes, which we then further converted into EUV fluxes in the 100 to 912\,\AA\ range employing the scaling relations of \citet{linsky2014}. We followed this procedure separately considering F5, G5, K5, and M0 stars. Figure~\ref{fig:fig6} shows a comparison between the EUV$-$\logR\ correlations obtained combining the scaling relations of \citet{linsky2013} and \citet{linsky2014} and the two linear fits derived in this work, including the uncertainties on the fits. 

We find that our fit for F-, G-, K-, M-type stars is significantly steeper than that obtained by combining the scaling relations of \citet{linsky2013} and \citet{linsky2014}. For active stars, the two correlations lead to comparable results, while, as a consequence of the different slopes, difference of up to about 1.5 orders of magnitude can be observed for inactive F-, G-, K-type stars and of the order of less than one magnitude in the case of inactive M-type stars. We also followed the original formalism described by \citet{noyes1984} to compute $F_{\rm HK}$(1 AU) from the \logR\ values obtaining no significant differences from the results described above.

\begin{figure}[h]
\begin{center}
\includegraphics[width=\hsize]{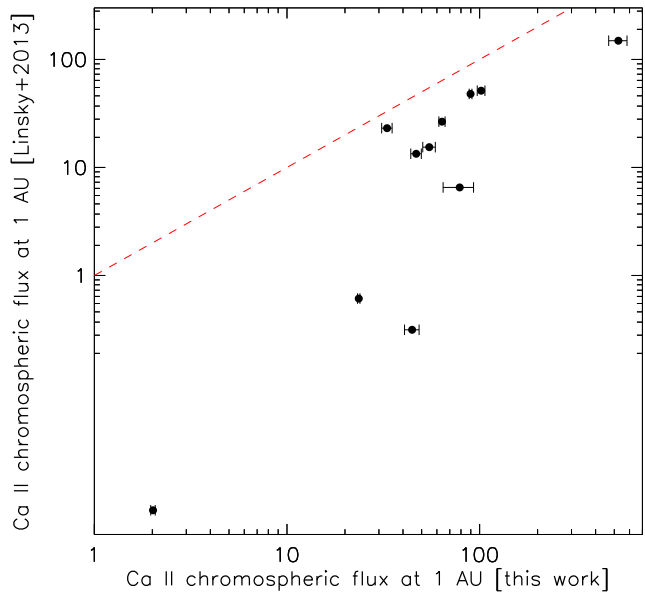}
\caption{Comparison between  Ca\,{\sc ii}\,H\&K chromospheric emission flux at 1 AU calculated from \logR\ employing the method described in section~\ref{sec:discussion} and that used in \citet{linsky2013}. The red dashed line represents the linear relation. We observe that the method described in Section~\ref{sec:discussion} over estimates the Ca\,{\sc ii}\,H\&K chromospheric emission flux as compared to those provided in \citet{linsky2013}.}
\label{fig:fig7}
\end{center}
\end{figure}

In an attempt to understand the origin of these differences, we took from \citet{linsky2014} the EUV fluxes of the stars that are in common with our sample and plot their position in Figure~\ref{fig:fig6}, obtaining that they follow our EUV vs \logR\ fits. To further trace the origin of this difference, we compared the Ca\,{\sc ii}\,H\&K chromospheric emission flux at 1\,AU calculated from \logR\ using the method described above and the values given by \citet[][their Table~4]{linsky2013}. The comparison is shown in Figure~\ref{fig:fig7}. Indeed, Equation~(\ref{eq:E}) appears to overestimate the Ca\,{\sc ii}\,H\&K chromospheric emission flux at 1\,AU, particularly for the less active stars. This further comparison suggests that the difference may be due to the fact that Equation~(\ref{eq:E}) does not account for the extra chromospheric emission flux falling outside of the band employed to measure $S_{\rm MW}$ values, and hence \logR\ values \citep{hartmann1984}. In other words, the wavelength bands considered for estimating the \logR\ values and the $F_{\rm HK}$(1 AU) values employed by \citet{linsky2013} are different. Therefore deriving $F_{\rm HK}$(1 AU) from the \logR\ value and then employing the \citet{linsky2013} and \citet{linsky2014} scaling relations may significantly overestimate the EUV fluxes, and hence our Equation~(\ref{eq:our_result}) shall be used, instead. Our results can be therefore used to estimate wavelength-integrated stellar EUV fluxes for stars for which X-ray or FUV observations are either not available or not possible, hence enabling one to more accurately study, for example, the upper atmospheres of planets orbiting late-type stars and their interaction with the host star. 
%

\begin{acknowledgements}
A.G.S. and L.F. acknowledge financial support from the Austrian Forschungsf\"orderungsgesellschaft FFG project CONTROL P865968. We thank the anonymous referee for the comments that led to improving the manuscript. This research has made use of the VizieR catalogue access tool, CDS, Strasbourg, France (DOI: 10.26093/cds/vizier). The original description of the VizieR service was published in 2000, A\&AS 143, 23.
\end{acknowledgements}


\bibliographystyle{aa}
\bibliography{reference.bib} 

\begin{thebibliography}{86}
\expandafter\ifx\csname natexlab\endcsname\relax\def\natexlab#1{#1}\fi

\bibitem[{{Anderson} {et~al.}(2014){Anderson}, {Collier Cameron}, {Delrez},
  {Doyle}, {Faedi}, {Fumel}, {Gillon}, {G{\'o}mez Maqueo Chew}, {Hellier},
  {Jehin}, {Lendl}, {Maxted}, {Pepe}, {Pollacco}, {Queloz}, {S{\'e}gransan},
  {Skillen}, {Smalley}, {Smith}, {Southworth}, {Triaud}, {Turner}, {Udry}, \&
  {West}}]{Anderson2014}
{Anderson}, D.~R., {Collier Cameron}, A., {Delrez}, L., {et~al.} 2014, \mnras,
  445, 1114

\bibitem[{{Arriagada}(2011)}]{Arriagada2011}
{Arriagada}, P. 2011, \apj, 734, 70

\bibitem[{{Astudillo-Defru} {et~al.}(2017){Astudillo-Defru}, {Delfosse},
  {Bonfils}, {Forveille}, {Lovis}, \& {Rameau}}]{Astudillo2017}
{Astudillo-Defru}, N., {Delfosse}, X., {Bonfils}, X., {et~al.} 2017, \aap, 600,
  A13

\bibitem[{{Baliunas} {et~al.}(1995){Baliunas}, {Donahue}, {Soon}, {Horne},
  {Frazer}, {Woodard-Eklund}, {Bradford}, {Rao}, {Wilson}, {Zhang}, {Bennett},
  {Briggs}, {Carroll}, {Duncan}, {Figueroa}, {Lanning}, {Misch}, {Mueller},
  {Noyes}, {Poppe}, {Porter}, {Robinson}, {Russell}, {Shelton}, {Soyumer},
  {Vaughan}, \& {Whitney}}]{Baliunas1995}
{Baliunas}, S.~L., {Donahue}, R.~A., {Soon}, W.~H., {et~al.} 1995, \apj, 438,
  269

\bibitem[{{Boro Saikia} {et~al.}(2018){Boro Saikia}, {Marvin}, {Jeffers},
  {Reiners}, {Cameron}, {Marsden}, {Petit}, {Warnecke}, \&
  {Yadav}}]{borosaikia2018}
{Boro Saikia}, S., {Marvin}, C.~J., {Jeffers}, S.~V., {et~al.} 2018, \aap, 616,
  A108

\bibitem[{{Bourrier} {et~al.}(2018){Bourrier}, {Lecavelier des Etangs},
  {Ehrenreich}, {Sanz-Forcada}, {Allart}, {Ballester}, {Buchhave}, {Cohen},
  {Deming}, {Evans}, {Garc{\'\i}a Mu{\~n}oz}, {Henry}, {Kataria}, {Lavvas},
  {Lewis}, {L{\'o}pez-Morales}, {Marley}, {Sing}, \& {Wakeford}}]{bourrier2018}
{Bourrier}, V., {Lecavelier des Etangs}, A., {Ehrenreich}, D., {et~al.} 2018,
  \aap, 620, A147

\bibitem[{{Caillault} {et~al.}(1991){Caillault}, {Vilhu}, \&
  {Linsky}}]{Caillault1991}
{Caillault}, J.-P., {Vilhu}, O., \& {Linsky}, J.~L. 1991, \apj, 383, 594

\bibitem[{{Canto Martins} {et~al.}(2011){Canto Martins}, {Das Chagas}, {Alves},
  {Le{\~a}o}, {de Souza Neto}, \& {de Medeiros}}]{CantoMartins2011}
{Canto Martins}, B.~L., {Das Chagas}, M.~L., {Alves}, S., {et~al.} 2011, \aap,
  530, A73

\bibitem[{{Cecchi-Pestellini} {et~al.}(2009){Cecchi-Pestellini}, {Ciaravella},
  {Micela}, \& {Penz}}]{cecchi2009}
{Cecchi-Pestellini}, C., {Ciaravella}, A., {Micela}, G., \& {Penz}, T. 2009,
  \aap, 496, 863

\bibitem[{{Chadney} {et~al.}(2015){Chadney}, {Galand}, {Unruh}, {Koskinen}, \&
  {Sanz-Forcada}}]{Chadney2015}
{Chadney}, J.~M., {Galand}, M., {Unruh}, Y.~C., {Koskinen}, T.~T., \&
  {Sanz-Forcada}, J. 2015, \icarus, 250, 357

\bibitem[{{Cubillos} {et~al.}(2017){Cubillos}, {Erkaev}, {Juvan}, {Fossati},
  {Johnstone}, {Lammer}, {Lendl}, {Odert}, \& {Kislyakova}}]{cubillos2017}
{Cubillos}, P., {Erkaev}, N.~V., {Juvan}, I., {et~al.} 2017, \mnras, 466, 1868

\bibitem[{{Cubillos} {et~al.}(2020){Cubillos}, {Fossati}, {Koskinen}, {Young},
  {Salz}, {France}, {Sreejith}, \& {Haswell}}]{cubillos2020}
{Cubillos}, P.~E., {Fossati}, L., {Koskinen}, T., {et~al.} 2020, \aj, 159, 111

\bibitem[{{Duncan} {et~al.}(1991){Duncan}, {Vaughan}, {Wilson}, {Preston},
  {Frazer}, {Lanning}, {Misch}, {Mueller}, {Soyumer}, {Woodard}, {Baliunas},
  {Noyes}, {Hartmann}, {Porter}, {Zwaan}, {Middelkoop}, {Rutten}, \&
  {Mihalas}}]{Duncan1991}
{Duncan}, D.~K., {Vaughan}, A.~H., {Wilson}, O.~C., {et~al.} 1991, \apjs, 76,
  383

\bibitem[{{Ehrenreich} {et~al.}(2015){Ehrenreich}, {Bourrier}, {Wheatley},
  {Lecavelier des Etangs}, {H{\'e}brard}, {Udry}, {Bonfils}, {Delfosse},
  {D{\'e}sert}, {Sing}, \& {Vidal-Madjar}}]{ehrenreich2015}
{Ehrenreich}, D., {Bourrier}, V., {Wheatley}, P.~J., {et~al.} 2015, \nat, 522,
  459

\bibitem[{{Ehrenreich} \& {D{\'e}sert}(2011)}]{ehrenreich2011}
{Ehrenreich}, D. \& {D{\'e}sert}, J.~M. 2011, \aap, 529, A136

\bibitem[{{Fossati} {et~al.}(2017{\natexlab{a}}){Fossati}, {Erkaev}, {Lammer},
  {Cubillos}, {Odert}, {Juvan}, {Kislyakova}, {Lendl}, {Kubyshkina}, \&
  {Bauer}}]{fossati2017}
{Fossati}, L., {Erkaev}, N.~V., {Lammer}, H., {et~al.} 2017{\natexlab{a}},
  \aap, 598, A90

\bibitem[{{Fossati} {et~al.}(2015){Fossati}, {France}, {Koskinen}, {Juvan},
  {Haswell}, \& {Lendl}}]{fossati2015}
{Fossati}, L., {France}, K., {Koskinen}, T., {et~al.} 2015, \apj, 815, 118

\bibitem[{{Fossati} {et~al.}(2010){Fossati}, {Haswell}, {Froning}, {Hebb},
  {Holmes}, {Kolb}, {Helling}, {Carter}, {Wheatley}, {Collier Cameron},
  {Loeillet}, {Pollacco}, {Street}, {Stempels}, {Simpson}, {Udry}, {Joshi},
  {West}, {Skillen}, \& {Wilson}}]{fossati2010}
{Fossati}, L., {Haswell}, C.~A., {Froning}, C.~S., {et~al.} 2010, \apjl, 714,
  L222

\bibitem[{{Fossati} {et~al.}(2018{\natexlab{a}}){Fossati}, {Koskinen},
  {France}, {Cubillos}, {Haswell}, {Lanza}, \& {Pillitteri}}]{fossati2018}
{Fossati}, L., {Koskinen}, T., {France}, K., {et~al.} 2018{\natexlab{a}}, \aj,
  155, 113

\bibitem[{{Fossati} {et~al.}(2018{\natexlab{b}}){Fossati}, {Koskinen},
  {Lothringer}, {France}, {Young}, \& {Sreejith}}]{fossati2018b}
{Fossati}, L., {Koskinen}, T., {Lothringer}, J.~D., {et~al.}
  2018{\natexlab{b}}, \apjl, 868, L30

\bibitem[{{Fossati} {et~al.}(2017{\natexlab{b}}){Fossati}, {Marcelja}, {Staab},
  {Cubillos}, {France}, {Haswell}, {Ingrassia}, {Jenkins}, {Koskinen}, {Lanza},
  {Redfield}, {Youngblood}, \& {Pelzmann}}]{fossati2017b}
{Fossati}, L., {Marcelja}, S.~E., {Staab}, D., {et~al.} 2017{\natexlab{b}},
  \aap, 601, A104

\bibitem[{{France} {et~al.}(2018){France}, {Arulanantham}, {Fossati}, {Lanza},
  {Loyd}, {Redfield}, \& {Schneider}}]{france2018}
{France}, K., {Arulanantham}, N., {Fossati}, L., {et~al.} 2018, \apjs, 239, 16

\bibitem[{{France} {et~al.}(2019){France}, {Fleming}, {Drake}, {Mason},
  {Youngblood}, {Bourrier}, {Fossati}, {Froning}, {Koskinen}, {Kruczek},
  {Lipscy}, {McEntaffer}, {Romaine}, {Siegmund}, \& {Wilkinson}}]{france2019}
{France}, K., {Fleming}, B.~T., {Drake}, J.~J., {et~al.} 2019, in Society of
  Photo-Optical Instrumentation Engineers (SPIE) Conference Series, Vol. 11118,
  \procspie, 1111808

\bibitem[{{Gaia Collaboration} {et~al.}(2018){Gaia Collaboration}, {Brown},
  {Vallenari}, {Prusti}, {de Bruijne}, {Babusiaux}, {Bailer-Jones}, {Biermann},
  {Evans}, {Eyer}, {Jansen}, {Jordi}, {Klioner}, {Lammers}, {Lindegren},
  {Luri}, {Mignard}, {Panem}, {Pourbaix}, {Randich}, {Sartoretti}, {Siddiqui},
  {Soubiran}, {van Leeuwen}, {Walton}, {Arenou}, {Bastian}, {Cropper},
  {Drimmel}, {Katz}, {Lattanzi}, {Bakker}, {Cacciari}, {Casta{\~n}eda},
  {Chaoul}, {Cheek}, {De Angeli}, {Fabricius}, {Guerra}, {Holl}, {Masana},
  {Messineo}, {Mowlavi}, {Nienartowicz}, {Panuzzo}, {Portell}, {Riello},
  {Seabroke}, {Tanga}, {Th{\'e}venin}, {Gracia-Abril}, {Comoretto},
  {Garcia-Reinaldos}, {Teyssier}, {Altmann}, {Andrae}, {Audard},
  {Bellas-Velidis}, {Benson}, {Berthier}, {Blomme}, {Burgess}, {Busso},
  {Carry}, {Cellino}, {Clementini}, {Clotet}, {Creevey}, {Davidson}, {De
  Ridder}, {Delchambre}, {Dell'Oro}, {Ducourant},
  {Fern{\'a}ndez-Hern{\'a}ndez}, {Fouesneau}, {Fr{\'e}mat}, {Galluccio},
  {Garc{\'\i}a-Torres}, {Gonz{\'a}lez-N{\'u}{\~n}ez}, {Gonz{\'a}lez-Vidal},
  {Gosset}, {Guy}, {Halbwachs}, {Hambly}, {Harrison}, {Hern{\'a}ndez},
  {Hestroffer}, {Hodgkin}, {Hutton}, {Jasniewicz}, {Jean-Antoine-Piccolo},
  {Jordan}, {Korn}, {Krone-Martins}, {Lanzafame}, {Lebzelter}, {L{\"o}ffler},
  {Manteiga}, {Marrese}, {Mart{\'\i}n-Fleitas}, {Moitinho}, {Mora}, {Muinonen},
  {Osinde}, {Pancino}, {Pauwels}, {Petit}, {Recio-Blanco}, {Richards},
  {Rimoldini}, {Robin}, {Sarro}, {Siopis}, {Smith}, {Sozzetti}, {S{\"u}veges},
  {Torra}, {van Reeven}, {Abbas}, {Abreu Aramburu}, {Accart}, {Aerts},
  {Altavilla}, {{\'A}lvarez}, {Alvarez}, {Alves}, {Anderson}, {Andrei},
  {Anglada Varela}, {Antiche}, {Antoja}, {Arcay}, {Astraatmadja}, {Bach},
  {Baker}, {Balaguer-N{\'u}{\~n}ez}, {Balm}, {Barache}, {Barata}, {Barbato},
  {Barblan}, {Barklem}, {Barrado}, {Barros}, {Barstow}, {Bartholom{\'e}
  Mu{\~n}oz}, {Bassilana}, {Becciani}, {Bellazzini}, {Berihuete}, {Bertone},
  {Bianchi}, {Bienaym{\'e}}, {Blanco-Cuaresma}, {Boch}, {Boeche}, {Bombrun},
  {Borrachero}, {Bossini}, {Bouquillon}, {Bourda}, {Bragaglia}, {Bramante},
  {Breddels}, {Bressan}, {Brouillet}, {Br{\"u}semeister}, {Brugaletta},
  {Bucciarelli}, {Burlacu}, {Busonero}, {Butkevich}, {Buzzi}, {Caffau},
  {Cancelliere}, {Cannizzaro}, {Cantat-Gaudin}, {Carballo}, {Carlucci},
  {Carrasco}, {Casamiquela}, {Castellani}, {Castro-Ginard}, {Charlot},
  {Chemin}, {Chiavassa}, {Cocozza}, {Costigan}, {Cowell}, {Crifo}, {Crosta},
  {Crowley}, {Cuypers}, {Dafonte}, {Damerdji}, {Dapergolas}, {David}, {David},
  {de Laverny}, {De Luise}, {De March}, {de Martino}, {de Souza}, {de Torres},
  {Debosscher}, {del Pozo}, {Delbo}, {Delgado}, {Delgado}, {Di Matteo},
  {Diakite}, {Diener}, {Distefano}, {Dolding}, {Drazinos}, {Dur{\'a}n},
  {Edvardsson}, {Enke}, {Eriksson}, {Esquej}, {Eynard Bontemps}, {Fabre},
  {Fabrizio}, {Faigler}, {Falc{\~a}o}, {Farr{\`a}s Casas}, {Federici},
  {Fedorets}, {Fernique}, {Figueras}, {Filippi}, {Findeisen}, {Fonti},
  {Fraile}, {Fraser}, {Fr{\'e}zouls}, {Gai}, {Galleti}, {Garabato},
  {Garc{\'\i}a-Sedano}, {Garofalo}, {Garralda}, {Gavel}, {Gavras}, {Gerssen},
  {Geyer}, {Giacobbe}, {Gilmore}, {Girona}, {Giuffrida}, {Glass}, {Gomes},
  {Granvik}, {Gueguen}, {Guerrier}, {Guiraud}, {Guti{\'e}rrez-S{\'a}nchez},
  {Haigron}, {Hatzidimitriou}, {Hauser}, {Haywood}, {Heiter}, {Helmi}, {Heu},
  {Hilger}, {Hobbs}, {Hofmann}, {Holland}, {Huckle}, {Hypki}, {Icardi},
  {Jan{\ss}en}, {Jevardat de Fombelle}, {Jonker}, {Juh{\'a}sz}, {Julbe},
  {Karampelas}, {Kewley}, {Klar}, {Kochoska}, {Kohley}, {Kolenberg},
  {Kontizas}, {Kontizas}, {Koposov}, {Kordopatis}, {Kostrzewa-Rutkowska},
  {Koubsky}, {Lambert}, {Lanza}, {Lasne}, {Lavigne}, {Le Fustec}, {Le
  Poncin-Lafitte}, {Lebreton}, {Leccia}, {Leclerc}, {Lecoeur-Taibi},
  {Lenhardt}, {Leroux}, {Liao}, {Licata}, {Lindstr{\o}m}, {Lister}, {Livanou},
  {Lobel}, {L{\'o}pez}, {Managau}, {Mann}, {Mantelet}, {Marchal}, {Marchant},
  {Marconi}, {Marinoni}, {Marschalk{\'o}}, {Marshall}, {Martino}, {Marton},
  {Mary}, {Massari}, {Matijevi{\v{c}}}, {Mazeh}, {McMillan}, {Messina},
  {Michalik}, {Millar}, {Molina}, {Molinaro}, {Moln{\'a}r}, {Montegriffo},
  {Mor}, {Morbidelli}, {Morel}, {Morris}, {Mulone}, {Muraveva}, {Musella},
  {Nelemans}, {Nicastro}, {Noval}, {O'Mullane}, {Ord{\'e}novic},
  {Ord{\'o}{\~n}ez-Blanco}, {Osborne}, {Pagani}, {Pagano}, {Pailler},
  {Palacin}, {Palaversa}, {Panahi}, {Pawlak}, {Piersimoni}, {Pineau}, {Plachy},
  {Plum}, {Poggio}, {Poujoulet}, {Pr{\v{s}}a}, {Pulone}, {Racero}, {Ragaini},
  {Rambaux}, {Ramos-Lerate}, {Regibo}, {Reyl{\'e}}, {Riclet}, {Ripepi}, {Riva},
  {Rivard}, {Rixon}, {Roegiers}, {Roelens}, {Romero-G{\'o}mez}, {Rowell},
  {Royer}, {Ruiz-Dern}, {Sadowski}, {Sagrist{\`a} Sell{\'e}s}, {Sahlmann},
  {Salgado}, {Salguero}, {Sanna}, {Santana-Ros}, {Sarasso}, {Savietto},
  {Schultheis}, {Sciacca}, {Segol}, {Segovia}, {S{\'e}gransan}, {Shih},
  {Siltala}, {Silva}, {Smart}, {Smith}, {Solano}, {Solitro}, {Sordo}, {Soria
  Nieto}, {Souchay}, {Spagna}, {Spoto}, {Stampa}, {Steele},
  {Steidelm{\"u}ller}, {Stephenson}, {Stoev}, {Suess}, {Surdej}, {Szabados},
  {Szegedi-Elek}, {Tapiador}, {Taris}, {Tauran}, {Taylor}, {Teixeira},
  {Terrett}, {Teyssand ier}, {Thuillot}, {Titarenko}, {Torra Clotet}, {Turon},
  {Ulla}, {Utrilla}, {Uzzi}, {Vaillant}, {Valentini}, {Valette}, {van Elteren},
  {Van Hemelryck}, {van Leeuwen}, {Vaschetto}, {Vecchiato}, {Veljanoski},
  {Viala}, {Vicente}, {Vogt}, {von Essen}, {Voss}, {Votruba}, {Voutsinas},
  {Walmsley}, {Weiler}, {Wertz}, {Wevers}, {Wyrzykowski}, {Yoldas},
  {{\v{Z}}erjal}, {Ziaeepour}, {Zorec}, {Zschocke}, {Zucker}, {Zurbach}, \&
  {Zwitter}}]{gaiadr2}
{Gaia Collaboration}, {Brown}, A.~G.~A., {Vallenari}, A., {et~al.} 2018, \aap,
  616, A1

\bibitem[{{Garc{\'\i}a Mu{\~n}oz} \& {Schneider}(2019)}]{garcia2019}
{Garc{\'\i}a Mu{\~n}oz}, A. \& {Schneider}, P.~C. 2019, \apjl, 884, L43

\bibitem[{{Gray} {et~al.}(2006){Gray}, {Corbally}, {Garrison}, {McFadden},
  {Bubar}, {McGahee}, {O'Donoghue}, \& {Knox}}]{gray2006}
{Gray}, R.~O., {Corbally}, C.~J., {Garrison}, R.~F., {et~al.} 2006, \aj, 132,
  161

\bibitem[{{Gray} {et~al.}(2003){Gray}, {Corbally}, {Garrison}, {McFadden}, \&
  {Robinson}}]{gray2003}
{Gray}, R.~O., {Corbally}, C.~J., {Garrison}, R.~F., {McFadden}, M.~T., \&
  {Robinson}, P.~E. 2003, \aj, 126, 2048

\bibitem[{Hall {et~al.}(2009)Hall, Henry, Lockwood, Skiff, \& Saar}]{Hall2009}
Hall, J.~C., Henry, G.~W., Lockwood, G.~W., Skiff, B.~A., \& Saar, S.~H. 2009,
  The Astronomical Journal, 138, 312

\bibitem[{{Hartmann} {et~al.}(1984){Hartmann}, {Soderblom}, {Noyes}, {Burnham},
  \& {Vaughan}}]{hartmann1984}
{Hartmann}, L., {Soderblom}, D.~R., {Noyes}, R.~W., {Burnham}, N., \&
  {Vaughan}, A.~H. 1984, \apj, 276, 254

\bibitem[{{Haswell} {et~al.}(2012){Haswell}, {Fossati}, {Ayres}, {France},
  {Froning}, {Holmes}, {Kolb}, {Busuttil}, {Street}, {Hebb}, {Collier Cameron},
  {Enoch}, {Burwitz}, {Rodriguez}, {West}, {Pollacco}, {Wheatley}, \&
  {Carter}}]{haswell2012}
{Haswell}, C.~A., {Fossati}, L., {Ayres}, T., {et~al.} 2012, \apj, 760, 79

\bibitem[{{Henry} {et~al.}(1996){Henry}, {Soderblom}, {Donahue}, \&
  {Baliunas}}]{henry1996}
{Henry}, T.~J., {Soderblom}, D.~R., {Donahue}, R.~A., \& {Baliunas}, S.~L.
  1996, \aj, 111, 439

\bibitem[{{Hinkel} {et~al.}(2017){Hinkel}, {Mamajek}, {Turnbull}, {Osby},
  {Shkolnik}, {Smith}, {Klimasewski}, {Somers}, \& {Desch}}]{hinkel2017}
{Hinkel}, N.~R., {Mamajek}, E.~E., {Turnbull}, M.~C., {et~al.} 2017, \apj, 848,
  34

\bibitem[{{Isaacson} \& {Fischer}(2010)}]{issacson2010}
{Isaacson}, H. \& {Fischer}, D. 2010, \apj, 725, 875

\bibitem[{{Jenkins} {et~al.}(2006){Jenkins}, {Jones}, {Tinney}, {Butler},
  {McCarthy}, {Marcy}, {Pinfield}, {Carter}, \& {Penny}}]{Jenkins2006}
{Jenkins}, J.~S., {Jones}, H.~R.~A., {Tinney}, C.~G., {et~al.} 2006, \mnras,
  372, 163

\bibitem[{{Jenkins} {et~al.}(2011){Jenkins}, {Murgas}, {Rojo}, {Jones},
  {Day-Jones}, {Jones}, {Clarke}, {Ruiz}, \& {Pinfield}}]{jenkins2011}
{Jenkins}, J.~S., {Murgas}, F., {Rojo}, P., {et~al.} 2011, \aap, 531, A8

\bibitem[{{Jin} \& {Mordasini}(2018)}]{jin2018}
{Jin}, S. \& {Mordasini}, C. 2018, \apj, 853, 163

\bibitem[{{Jin} {et~al.}(2014){Jin}, {Mordasini}, {Parmentier}, {van Boekel},
  {Henning}, \& {Ji}}]{jin2014}
{Jin}, S., {Mordasini}, C., {Parmentier}, V., {et~al.} 2014, \apj, 795, 65

\bibitem[{{Johnstone} {et~al.}(2015){Johnstone}, {G{\"u}del}, {Brott}, \&
  {L{\"u}ftinger}}]{johnstone2015}
{Johnstone}, C.~P., {G{\"u}del}, M., {Brott}, I., \& {L{\"u}ftinger}, T. 2015,
  \aap, 577, A28

\bibitem[{{Koskinen} {et~al.}(2013{\natexlab{a}}){Koskinen}, {Harris}, {Yelle},
  \& {Lavvas}}]{koskinen2013a}
{Koskinen}, T.~T., {Harris}, M.~J., {Yelle}, R.~V., \& {Lavvas}, P.
  2013{\natexlab{a}}, \icarus, 226, 1678

\bibitem[{{Koskinen} {et~al.}(2013{\natexlab{b}}){Koskinen}, {Yelle}, {Harris},
  \& {Lavvas}}]{koskinen2013b}
{Koskinen}, T.~T., {Yelle}, R.~V., {Harris}, M.~J., \& {Lavvas}, P.
  2013{\natexlab{b}}, \icarus, 226, 1695

\bibitem[{{Kubyshkina} {et~al.}(2019{\natexlab{a}}){Kubyshkina}, {Cubillos},
  {Fossati}, {Erkaev}, {Johnstone}, {Kislyakova}, {Lammer}, {Lendl}, {Odert},
  \& {G{\"u}del}}]{Kubyshkina2019a}
{Kubyshkina}, D., {Cubillos}, P.~E., {Fossati}, L., {et~al.}
  2019{\natexlab{a}}, \apj, 879, 26

\bibitem[{{Kubyshkina} {et~al.}(2018{\natexlab{a}}){Kubyshkina}, {Fossati},
  {Erkaev}, {Johnstone}, {Cubillos}, {Kislyakova}, {Lammer}, {Lendl}, \&
  {Odert}}]{kubyshkina2018}
{Kubyshkina}, D., {Fossati}, L., {Erkaev}, N.~V., {et~al.} 2018{\natexlab{a}},
  \aap, 619, A151

\bibitem[{{Kubyshkina} {et~al.}(2019{\natexlab{b}}){Kubyshkina}, {Fossati},
  {Mustill}, {Cubillos}, {Davies}, {Erkaev}, {Johnstone}, {Kislyakova},
  {Lammer}, {Lendl}, \& {Odert}}]{Kubyshkina2019b}
{Kubyshkina}, D., {Fossati}, L., {Mustill}, A.~J., {et~al.} 2019{\natexlab{b}},
  \aap, 632, A65

\bibitem[{{Kubyshkina} {et~al.}(2018{\natexlab{b}}){Kubyshkina}, {Lendl},
  {Fossati}, {Cubillos}, {Lammer}, {Erkaev}, \& {Johnstone}}]{kubyshkina2018a}
{Kubyshkina}, D., {Lendl}, M., {Fossati}, L., {et~al.} 2018{\natexlab{b}},
  \aap, 612, A25

\bibitem[{{Lammer} {et~al.}(2016){Lammer}, {Erkaev}, {Fossati}, {Juvan},
  {Odert}, {Cubillos}, {Guenther}, {Kislyakova}, {Johnstone}, {L{\"u}ftinger},
  \& {G{\"u}del}}]{lammer2016}
{Lammer}, H., {Erkaev}, N.~V., {Fossati}, L., {et~al.} 2016, \mnras, 461, L62

\bibitem[{{Lammer} {et~al.}(2020){Lammer}, {Leitzinger}, {Scherf}, {Odert},
  {Burger}, {Kubyshkina}, {Johnstone}, {Maindl}, {Sch{\"a}fer}, {G{\"u}del},
  {Tosi}, {Nikolaou}, {Marcq}, {Erkaev}, {Noack}, {Kislyakova}, {Fossati},
  {Pilat-Lohinger}, {Ragossnig}, \& {Dorfi}}]{lammer2020}
{Lammer}, H., {Leitzinger}, M., {Scherf}, M., {et~al.} 2020, \icarus, 339,
  113551

\bibitem[{{Lammer} {et~al.}(2003){Lammer}, {Selsis}, {Ribas}, {Guinan},
  {Bauer}, \& {Weiss}}]{lammer2003}
{Lammer}, H., {Selsis}, F., {Ribas}, I., {et~al.} 2003, \apjl, 598, L121

\bibitem[{{Lammer} {et~al.}(2018){Lammer}, {Zerkle}, {Gebauer}, {Tosi},
  {Noack}, {Scherf}, {Pilat-Lohinger}, {G{\"u}del}, {Grenfell}, {Godolt}, \&
  {Nikolaou}}]{lammer2018}
{Lammer}, H., {Zerkle}, A.~L., {Gebauer}, S., {et~al.} 2018, \aapr, 26, 2

\bibitem[{{Laws} {et~al.}(2003){Laws}, {Gonzalez}, {Walker}, {Tyagi},
  {Dodsworth}, {Snider}, \& {Suntzeff}}]{Laws2003}
{Laws}, C., {Gonzalez}, G., {Walker}, K.~M., {et~al.} 2003, \aj, 125, 2664

\bibitem[{{Lecavelier des Etangs} {et~al.}(2012){Lecavelier des Etangs},
  {Bourrier}, {Wheatley}, {Dupuy}, {Ehrenreich}, {Vidal-Madjar}, {H{\'e}brard},
  {Ballester}, {D{\'e}sert}, {Ferlet}, \& {Sing}}]{lecavelier2012}
{Lecavelier des Etangs}, A., {Bourrier}, V., {Wheatley}, P.~J., {et~al.} 2012,
  \aap, 543, L4

\bibitem[{{Linsky} {et~al.}(2014){Linsky}, {Fontenla}, \&
  {France}}]{linsky2014}
{Linsky}, J.~L., {Fontenla}, J., \& {France}, K. 2014, \apj, 780, 61

\bibitem[{{Linsky} {et~al.}(2013){Linsky}, {France}, \& {Ayres}}]{linsky2013}
{Linsky}, J.~L., {France}, K., \& {Ayres}, T. 2013, \apj, 766, 69

\bibitem[{{Linsky} {et~al.}(2010){Linsky}, {Yang}, {France}, {Froning},
  {Green}, {Stocke}, \& {Osterman}}]{linsky2010}
{Linsky}, J.~L., {Yang}, H., {France}, K., {et~al.} 2010, \apj, 717, 1291

\bibitem[{{Lopez} \& {Fortney}(2013)}]{lopez2013}
{Lopez}, E.~D. \& {Fortney}, J.~J. 2013, \apj, 776, 2

\bibitem[{{Louden} {et~al.}(2017){Louden}, {Wheatley}, \&
  {Briggs}}]{louden2017}
{Louden}, T., {Wheatley}, P.~J., \& {Briggs}, K. 2017, \mnras, 464, 2396

\bibitem[{{Mamajek} \& {Hillenbrand}(2008)}]{Mamajek2008}
{Mamajek}, E.~E. \& {Hillenbrand}, L.~A. 2008, \apj, 687, 1264

\bibitem[{{Mansfield} {et~al.}(2018){Mansfield}, {Bean}, {Oklop{\v{c}}i{\'c}},
  {Kreidberg}, {D{\'e}sert}, {Kempton}, {Line}, {Fortney}, {Henry}, {Mallonn},
  {Stevenson}, {Dragomir}, {Allart}, \& {Bourrier}}]{mansfield2018}
{Mansfield}, M., {Bean}, J.~L., {Oklop{\v{c}}i{\'c}}, A., {et~al.} 2018, \apjl,
  868, L34

\bibitem[{{Ment} {et~al.}(2018){Ment}, {Fischer}, {Bakos}, {Howard}, \&
  {Isaacson}}]{ment2018}
{Ment}, K., {Fischer}, D.~A., {Bakos}, G., {Howard}, A.~W., \& {Isaacson}, H.
  2018, \aj, 156, 213

\bibitem[{{Mitani} {et~al.}(2020){Mitani}, {Nakatani}, \&
  {Yoshida}}]{hiroto2020}
{Mitani}, H., {Nakatani}, R., \& {Yoshida}, N. 2020, arXiv e-prints,
  arXiv:2005.08676

\bibitem[{{Mittag} {et~al.}(2013){Mittag}, {Schmitt}, \&
  {Schr{\"o}der}}]{mittag2013}
{Mittag}, M., {Schmitt}, J.~H.~M.~M., \& {Schr{\"o}der}, K.~P. 2013, \aap, 549,
  A117

\bibitem[{Morris {et~al.}(2017)Morris, Hawley, Hebb, Sakari, Davenport,
  Isaacson, Howard, Montet, \& Agol}]{Morris2017}
Morris, B.~M., Hawley, S.~L., Hebb, L., {et~al.} 2017, The Astrophysical
  Journal, 848, 58

\bibitem[{{Moutou} {et~al.}(2014){Moutou}, {H{\'e}brard}, {Bouchy}, {Arnold},
  {Santos}, {Astudillo-Defru}, {Boisse}, {Bonfils}, {Borgniet}, {Delfosse},
  {D{\'\i}az}, {Ehrenreich}, {Forveille}, {Gregorio}, {Labrevoir}, {Lagrange},
  {Montagnier}, {Montalto}, {Pepe}, {Sahlmann}, {Santerne}, {S{\'e}gransan},
  {Udry}, \& {Vanhuysse}}]{Moutou2014}
{Moutou}, C., {H{\'e}brard}, G., {Bouchy}, F., {et~al.} 2014, \aap, 563, A22

\bibitem[{{Murray-Clay} {et~al.}(2009){Murray-Clay}, {Chiang}, \&
  {Murray}}]{murray2009}
{Murray-Clay}, R.~A., {Chiang}, E.~I., \& {Murray}, N. 2009, \apj, 693, 23

\bibitem[{{Noyes} {et~al.}(1984){Noyes}, {Hartmann}, {Baliunas}, {Duncan}, \&
  {Vaughan}}]{noyes1984}
{Noyes}, R.~W., {Hartmann}, L.~W., {Baliunas}, S.~L., {Duncan}, D.~K., \&
  {Vaughan}, A.~H. 1984, \apj, 279, 763

\bibitem[{{Owen} \& {Wu}(2013)}]{owen2013}
{Owen}, J.~E. \& {Wu}, Y. 2013, \apj, 775, 105

\bibitem[{{Owen} \& {Wu}(2016)}]{owen2016}
{Owen}, J.~E. \& {Wu}, Y. 2016, \apj, 817, 107

\bibitem[{{Owen} \& {Wu}(2017)}]{owen2017}
{Owen}, J.~E. \& {Wu}, Y. 2017, \apj, 847, 29

\bibitem[{{Pace}(2013)}]{pace2013}
{Pace}, G. 2013, \aap, 551, L8

\bibitem[{{Pecaut} \& {Mamajek}(2013)}]{pecaut2013}
{Pecaut}, M.~J. \& {Mamajek}, E.~E. 2013, \apjs, 208, 9

\bibitem[{{Penz} {et~al.}(2008){Penz}, {Micela}, \& {Lammer}}]{penz2008}
{Penz}, T., {Micela}, G., \& {Lammer}, H. 2008, \aap, 477, 309

\bibitem[{{Ribas} {et~al.}(2005){Ribas}, {Guinan}, {G{\"u}del}, \&
  {Audard}}]{ribas2005}
{Ribas}, I., {Guinan}, E.~F., {G{\"u}del}, M., \& {Audard}, M. 2005, \apj, 622,
  680

\bibitem[{{Robertson} {et~al.}(2012){Robertson}, {Endl}, {Cochran}, {MacQueen},
  {Wittenmyer}, {Horner}, {Brugamyer}, {Simon}, {Barnes}, \&
  {Caldwell}}]{Robertson2012}
{Robertson}, P., {Endl}, M., {Cochran}, W.~D., {et~al.} 2012, \apj, 749, 39

\bibitem[{{Rutten}(1984)}]{rutten1984}
{Rutten}, R.~G.~M. 1984, \aap, 130, 353

\bibitem[{{Salz} {et~al.}(2016){Salz}, {Czesla}, {Schneider}, \&
  {Schmitt}}]{salz2016}
{Salz}, M., {Czesla}, S., {Schneider}, P.~C., \& {Schmitt}, J.~H.~M.~M. 2016,
  \aap, 586, A75

\bibitem[{{Sanz-Forcada} {et~al.}(2011){Sanz-Forcada}, {Micela}, {Ribas},
  {Pollock}, {Eiroa}, {Velasco}, {Solano}, \&
  {Garc{\'\i}a-{\'A}lvarez}}]{Sanz-Forcada2011}
{Sanz-Forcada}, J., {Micela}, G., {Ribas}, I., {et~al.} 2011, \aap, 532, A6

\bibitem[{{Sing} {et~al.}(2019){Sing}, {Lavvas}, {Ballester}, {Lecavelier des
  Etangs}, {Marley}, {Nikolov}, {Ben-Jaffel}, {Bourrier}, {Buchhave}, {Deming},
  {Ehrenreich}, {Mikal-Evans}, {Kataria}, {Lewis}, {L{\'o}pez-Morales},
  {Garc{\'\i}a Mu{\~n}oz}, {Henry}, {Sanz-Forcada}, {Spake}, {Wakeford}, \&
  {PanCET Collaboration}}]{sing2019}
{Sing}, D.~K., {Lavvas}, P., {Ballester}, G.~E., {et~al.} 2019, \aj, 158, 91

\bibitem[{{Sreejith} {et~al.}(2019){Sreejith}, {Fossati}, {Fleming}, {France},
  {Koskinen}, {Egan}, {R{\"u}disser}, \& {Steller}}]{sreejith2019}
{Sreejith}, A.~G., {Fossati}, L., {Fleming}, B.~T., {et~al.} 2019, Journal of
  Astronomical Telescopes, Instruments, and Systems, 5, 018004

\bibitem[{{Tu} {et~al.}(2015){Tu}, {Johnstone}, {G{\"u}del}, \&
  {Lammer}}]{tu2015}
{Tu}, L., {Johnstone}, C.~P., {G{\"u}del}, M., \& {Lammer}, H. 2015, \aap, 577,
  L3

\bibitem[{{Vaughan} \& {Preston}(1980)}]{vaughan1980}
{Vaughan}, A.~H. \& {Preston}, G.~W. 1980, \pasp, 92, 385

\bibitem[{{Vidal-Madjar} {et~al.}(2003){Vidal-Madjar}, {Lecavelier des Etangs},
  {D{\'e}sert}, {Ballester}, {Ferlet}, {H{\'e}brard}, \& {Mayor}}]{vidal2003}
{Vidal-Madjar}, A., {Lecavelier des Etangs}, A., {D{\'e}sert}, J.~M., {et~al.}
  2003, \nat, 422, 143

\bibitem[{{Vidotto} {et~al.}(2018){Vidotto}, {Lichtenegger}, {Fossati},
  {Folsom}, {Wood}, {Murthy}, {Petit}, {Sreejith}, \& {Valyavin}}]{vidotto2018}
{Vidotto}, A.~A., {Lichtenegger}, H., {Fossati}, L., {et~al.} 2018, \mnras,
  481, 5296

\bibitem[{{Watson} {et~al.}(1981){Watson}, {Donahue}, \& {Walker}}]{watson1981}
{Watson}, A.~J., {Donahue}, T.~M., \& {Walker}, J.~C.~G. 1981, \icarus, 48, 150

\bibitem[{{Wilson}(1978)}]{wilson1978}
{Wilson}, O.~C. 1978, \apj, 226, 379

\bibitem[{{Wittenmyer} {et~al.}(2009){Wittenmyer}, {Endl}, {Cochran},
  {Levison}, \& {Henry}}]{Wittenmyer2009}
{Wittenmyer}, R.~A., {Endl}, M., {Cochran}, W.~D., {Levison}, H.~F., \&
  {Henry}, G.~W. 2009, \apjs, 182, 97

\bibitem[{{Wright} {et~al.}(2004){Wright}, {Marcy}, {Butler}, \&
  {Vogt}}]{wright2004}
{Wright}, J.~T., {Marcy}, G.~W., {Butler}, R.~P., \& {Vogt}, S.~S. 2004, \apjs,
  152, 261

\bibitem[{{Yelle}(2004)}]{yelle2004}
{Yelle}, R.~V. 2004, \icarus, 170, 167

\end{thebibliography}
\longtab{
\begin{landscape}
\begin{longtable}{l|cccccc|cccc|ccc}
\caption{\label{table2} List of selected targets in order of spectral type from F- to M-type, their basic parameters, \logR\ values, and EUV fluxes. The stellar bolometric fluxes (F$_{\rm bol}$) are at Earth.}\\ 
\hline\\
Name & SpT & V & B - V & T$_{\rm eff}$ & R$_*$ & d & \multicolumn{4}{c|}{log~$R'_{\rm HK}$} & F$_{\rm bol}$ & EUV (90$-$911\AA) & log (EUV/F$_{\rm bol}$) \\ 
&  & mag & mag & K & R$_{\rm sun}$ & pc & mean  & median & stddev & \# & 10$^{-7}$~erg&10$^{-14}$~erg & \\
&  &  &  &     &           &      &       &        &        & & ~s$^{-1}$~cm$^{-2}$ &~s$^{-1}$~cm$^{-2}$ & \\
\hline
\endhead
\hline \\
\endfoot
        HD 28568 &  F2V & 6.48 & 0.44 & 6502 & 1.68 & 45.21 & -4.39 & -4.39 & 0.038 & 1 & 0.708 & 286.83 & -4.39 $\pm$0.89 \\ 
        Procyon &  F5V & 0.37 & 0.42 & 6530 & 2.15 & 3.50 & -4.69 & -4.69 & 0.083 & 2 & 197.432 & 5495.00 & -5.56 $\pm$3.47 \\ 
        HD 120136 &  F7V & 4.49 & 0.49 & 6310 & 1.54 & 15.60 & -4.72 & -4.71 & 0.055 & 8 & 4.440 & 1015.14 & -4.64 $\pm$0.90 \\ 
        HD 197037 &  F7V & 6.81 & 0.50 & 6259 & 1.14 & 33.23 & -4.53 & -4.53 & 0.038 & 1 & 0.523 & 4.99 & -6.02 $\pm$1.01 \\ 
        HD 136118 &  F7V & 6.94 & 0.52 & 6148 & 1.73 & 51.51 & -4.89 & -4.88 & 0.065 & 5 & 0.465 & 8.73 & -5.73 $\pm$0.99 \\ 
        HD 9826 &  F9V & 4.10 & 0.54 & 6105 & 1.69 & 13.41 & -4.96 & -4.97 & 0.053 & 6 & 6.359 & 249.42 & -5.41 $\pm$0.96 \\ 
        HD 10647 &  F9V & 5.52 & 0.55 & 6143 & 1.12 & 17.34 & -4.66 & -4.67 & 0.048 & 3 & 1.720 & 72.33 & -5.38 $\pm$0.96 \\ 
        HD 23079 &  F9V & 7.11 & 0.57 & 5964 & 1.11 & 33.49 & -4.87 & -4.87 & 0.053 & 4 & 0.398 & 6.24 & -5.80 $\pm$1.00 \\ 
        HD 33262 &  F9V & 4.71 & 0.51 & 6227 & 1.06 & 11.62 & -4.35 & -4.35 & 0.038 & 2 & 3.633 & 586.13 & -4.79 $\pm$0.91 \\ 
        HD 106516 &  F9V & 6.11 & 0.46 & 6326 & 1.04 & 22.40 & -4.39 & -4.39 & 0.187 & 4 & 0.999 & 49.88 & -5.30 $\pm$1.01 \\ 
        HD 155358 &  G0V & 7.28 & 0.55 & 5987 & 1.32 & 43.67 & -4.88 & -4.88 & 0.075 & 2 & 0.340 & 2.49 & -6.13 $\pm$1.04 \\ 
        rho CrB &  G0V & 5.39 & 0.61 & 5853 & 1.32 & 17.48 & -5.01 & -4.99 & 0.059 & 7 & 1.938 & 14.97 & -6.11 $\pm$1.02 \\ 
        HD 39091 &  G0V & 5.67 & 0.58 & 5961 & 1.17 & 18.28 & -4.91 & -4.91 & 0.050 & 4 & 1.498 & 29.93 & -5.70 $\pm$0.98 \\ 
        HD 187085 &  G0V & 7.21 & 0.57 & 6026 & 1.42 & 45.96 & -4.89 & -4.89 & 0.060 & 2 & 0.363 & 6.24 & -5.76 $\pm$0.99 \\ 
        HD 209458 &  G0V & 7.63 & 0.58 & 6077 & 1.21 & 48.37 & -4.92 & -4.92 & 0.029 & 4 & 0.246 & 8.73 & -5.45 $\pm$1.03 \\ 
        HD 114729 A &  G0V & 6.69 & 0.62 & 5835 & 1.58 & 37.84 & -4.97 & -4.97 & 0.089 & 8 & 0.585 & 6.24 & -5.97 $\pm$1.01 \\ 
        HD 13931 &  G0V & 7.60 & 0.64 & 5838 & 1.31 & 47.46 & -4.98 & -4.98 & 0.044 & 4 & 0.253 & 8.73 & -5.46 $\pm$0.97 \\ 
        HD 28205 &  G0V & 7.40 & 0.55 & 6148 & 1.30 & 47.79 & -4.58 & -4.58 & 0.038 & 1 & 0.303 & 99.77 & -4.48 $\pm$0.90 \\ 
        HD 25825 &  G0V & 7.81 & 0.60 & 5965 & 1.14 & 47.56 & -4.42 & -4.42 & 0.060 & 4 & 0.208 & 132.19 & -4.20 $\pm$0.87 \\ 
        HD 97334 &  G0V & 6.41 & 0.61 & 5885 & 1.06 & 22.65 & -4.38 & -4.38 & 0.019 & 4 & 0.758 & 187.06 & -4.61 $\pm$0.90 \\ 
        HD 39587 &  G0V & 4.40 & 0.60 & 6028 & 1.00 & 8.84 & -4.42 & -4.42 & 0.024 & 6 & 4.824 & 1067.60 & -4.66 $\pm$2.98 \\ 
        16 Cyg A &  G1.5V & 5.95 & 0.64 & 5793 & 1.26 & 21.15 & -5.00 & -5.00 & 0.029 & 10 & 1.157 & 29.63 & -5.59 $\pm$1.23 \\ 
        HD 72905 &  G1.5V & 5.64 & 0.62 & 5839 & 0.98 & 14.45 & -4.34 & -4.35 & 0.045 & 6 & 1.540 & 1031.04 & -4.17 $\pm$1.12 \\ 
        HD 129333 &  G1.5V & 7.61 & 0.59 & 5584 & 1.03 & 34.45 & -4.10 & -4.11 & 0.062 & 7 & 0.251 & 1285.21 & -3.29 $\pm$1.06 \\ 
        47 UMa &  G1V & 5.04 & 0.62 & 5947 & 1.19 & 13.80 & -4.97 & -4.98 & 0.062 & 6 & 2.676 & 42.40 & -5.80 $\pm$0.99 \\ 
        HD 10180 &  G1V & 7.32 & 0.63 & 5891 & 1.20 & 39.00 & -4.92 & -4.95 & 0.092 & 5 & 0.328 & 8.73 & -5.57 $\pm$0.98 \\ 
        HD 117618 &  G2V & 7.17 & 0.60 & 5974 & 1.21 & 37.82 & -4.87 & -4.87 & 0.044 & 4 & 0.376 & 6.24 & -5.78 $\pm$0.99 \\ 
        HD 121504 &  G2V & 7.54 & 0.59 & 6004 & 1.11 & 41.71 & -4.78 & -4.80 & 0.086 & 5 & 0.268 & 13.72 & -5.29 $\pm$0.95 \\ 
        HD 199288 &  G2V & 6.52 & 0.59 & 5884 & 0.96 & 21.59 & -4.80 & -4.80 & 0.038 & 5 & 0.685 & 3.87 & -6.25 $\pm$1.03 \\ 
        alpha Cen A &  G2V & 0.01 & 0.71 & 5770 & 1.21 & 1.30 & -5.00 & -5.00 & 0.043 & 8 & 275.053 & 4553.00 & -5.78 $\pm$0.97 \\ 
        mu Ara &  G3V & 5.15 & 0.70 & 5805 & 1.34 & 15.61 & -5.01 & -5.02 & 0.039 & 3 & 2.418 & 44.90 & -5.73 $\pm$0.99 \\ 
        16 Cyg B &  G3V & 6.20 & 0.66 & 5777 & 1.13 & 21.15 & -5.00 & -5.00 & 0.029 & 10 & 0.919 & 27.34 & -5.53 $\pm$1.22 \\ 
        HD 1461 &  G3V & 6.60 & 0.67 & 5765 & 1.05 & 23.47 & -5.00 & -5.00 & 0.034 & 7 & 0.636 & 31.18 & -5.31 $\pm$0.95 \\ 
        HD 59967 &  G3V & 6.63 & 0.64 & 5806 & 0.94 & 21.77 & -4.36 & -4.36 & 0.017 & 5 & 0.616 & 224.48 & -4.44 $\pm$0.89 \\ 
        HD 37124 &  G4IV-V & 7.68 & 0.67 & 5677 & 0.89 & 31.69 & -4.88 & -4.90 & 0.029 & 5 & 0.235 & 2.49 & -5.97 $\pm$1.03 \\ 
        HD 38529 &  G4V & 5.92 & 0.77 & 5586 & 2.76 & 42.41 & -4.99 & -5.01 & 0.028 & 3 & 1.185 & 59.86 & -5.30 $\pm$0.95 \\ 
        HD 147513 &  G5V & 5.38 & 0.64 & 5883 & 0.97 & 12.91 & -4.57 & -4.57 & 0.082 & 4 & 1.963 & 158.38 & -5.09 $\pm$0.94 \\ 
        HD 222582 &  G5V & 7.69 & 0.65 & 5793 & 1.13 & 42.21 & -4.94 & -4.93 & 0.050 & 4 & 0.233 & 4.99 & -5.67 $\pm$0.99 \\ 
        HD 28185 &  G5V & 7.81 & 0.71 & 5665 & 1.05 & 39.43 & -5.02 & -5.02 & 0.038 & 6 & 0.209 & 8.73 & -5.38 $\pm$0.96 \\ 
        HD 4113 &  G5V & 7.88 & 0.73 & 5676 & 1.07 & 41.92 & -5.02 & -5.01 & 0.057 & 4 & 0.196 & 7.48 & -5.42 $\pm$0.96 \\ 
        HD 65216 &  G5V & 7.96 & 0.69 & 5658 & 0.87 & 35.16 & -4.90 & -4.90 & 0.034 & 2 & 0.182 & 7.48 & -5.39 $\pm$0.96 \\ 
        HD 178911 B &  G5V & 7.98 & 0.73 & 5596 & 1.03 & 41.02 & -4.90 & -4.90 & 0.060 & 7 & 0.178 & 13.72 & -5.11 $\pm$0.94 \\ 
        HD 79498 &  G5V & 8.02 & 0.71 & 5726 & 1.15 & 49.02 & -5.08 & -5.08 & 0.038 & 1 & 0.172 & 4.99 & -5.54 $\pm$0.98 \\ 
        HIP 91258 &  G5V & 8.65 & 0.01 & 5520 & 0.87 & 45.95 & -4.65 & -4.65 & 0.038 & 1 & 0.096 & 12.47 & -4.89 $\pm$0.93 \\ 
        HD 90156 &  G5V & 6.92 & 0.68 & 5705 & 0.87 & 21.96 & -4.93 & -4.94 & 0.032 & 8 & 0.474 & 6.24 & -5.88 $\pm$1.00 \\ 
        HD 20630 &  G5V & 4.85 & 0.67 & 5749 & 0.92 & 9.15 & -4.39 & -4.39 & 0.008 & 2 & 3.187 & 62.80 & -5.71 $\pm$7.08 \\ 
        HD 115617 &  G6.5V & 4.74 & 0.70 & 5915 & 0.85 & 8.51 & -4.97 & -4.96 & 0.037 & 9 & 3.527 & 49.88 & -5.85 $\pm$1.02 \\ 
        HD 43162 &  G6.5V & 6.37 & 0.70 & 5626 & 0.87 & 16.73 & -4.45 & -4.46 & 0.031 & 5 & 0.789 & 37.41 & -5.32 $\pm$1.00 \\ 
        HD 70642 &  G6V & 7.17 & 0.69 & 5667 & 1.04 & 29.30 & -4.91 & -4.90 & 0.045 & 5 & 0.376 & 17.46 & -5.33 $\pm$0.95 \\ 
        HD 47186 &  G6V & 7.63 & 0.73 & 5685 & 1.07 & 37.50 & -4.97 & -4.95 & 0.049 & 6 & 0.246 & 9.98 & -5.39 $\pm$0.96 \\ 
        HD 92788 &  G6V & 7.30 & 0.69 & 5722 & 1.14 & 34.69 & -5.01 & -5.01 & 0.041 & 5 & 0.334 & 13.72 & -5.39 $\pm$0.96 \\ 
        HD 102117 &  G6V & 7.47 & 0.72 & 5630 & 1.24 & 39.62 & -5.02 & -5.02 & 0.016 & 2 & 0.285 & 7.48 & -5.58 $\pm$0.98 \\ 
        HD 4208 &  G7V & 7.78 & 0.66 & 5644 & 0.93 & 34.23 & -4.89 & -4.89 & 0.046 & 9 & 0.214 & 3.74 & -5.76 $\pm$1.00 \\ 
        HD 10700 &  G8V & 3.50 & 0.72 & 5375 & 0.77 & 3.60 & -4.98 & -4.98 & 0.033 & 10 & 11.051 & 148.40 & -5.87 $\pm$1.00 \\ 
        HD 69830 &  G8V & 5.95 & 0.79 & 5500 & 0.83 & 12.56 & -4.99 & -4.99 & 0.004 & 5 & 1.157 & 19.95 & -5.76 $\pm$0.99 \\ 
        55 Cnc &  G8V & 5.95 & 0.87 & 5306 & 0.90 & 12.59 & -5.03 & -5.02 & 0.041 & 5 & 1.157 & 37.91 & -5.48 $\pm$0.97 \\ 
        HD 1237 &  G8V & 6.58 & 0.76 & 5520 & 0.87 & 17.56 & -4.40 & -4.41 & 0.081 & 7 & 0.649 & 259.40 & -4.40 $\pm$0.88 \\ 
        HD 154345 &  G8V & 6.74 & 0.76 & 5540 & 0.83 & 18.29 & -4.90 & -4.91 & 0.041 & 5 & 0.559 & 23.69 & -5.37 $\pm$0.96 \\ 
        HD 131156 &  G8V & 4.59 & 0.78 & 5550 & 0.81 & 6.70 & -4.32 & -4.32 & 0.018 & 5 & 4.038 & 1303.10 & -4.49 $\pm$0.61 \\ 
        GJ 86 &  G9V & 6.17 & 0.77 & 5350 & 0.68 & 10.79 & -4.76 & -4.76 & 0.019 & 4 & 0.945 & 54.87 & -5.24 $\pm$0.95 \\ 
        HD 147018 &  G9V & 8.30 & 0.76 & 5475 & 0.92 & 40.47 & -4.80 & -4.80 & 0.003 & 2 & 0.133 & 7.48 & -5.25 $\pm$0.95 \\ 
        HD 164922 &  G9V & 7.01 & 0.80 & 5342 & 0.95 & 22.02 & -5.07 & -5.07 & 0.015 & 6 & 0.436 & 7.48 & -5.77 $\pm$0.99 \\ 
        HD 7924 &  K0.5V & 7.18 & 0.83 & 5244 & 0.70 & 17.00 & -4.87 & -4.86 & 0.034 & 5 & 0.371 & 18.71 & -5.30 $\pm$0.95 \\ 
        HD 3651 &  K0.5V & 5.88 & 0.83 & 5280 & 0.83 & 11.14 & -5.04 & -5.05 & 0.035 & 6 & 1.234 & 47.39 & -5.42 $\pm$0.96 \\ 
        HD 166 &  K0V & 6.13 & 0.75 & 5552 & 0.82 & 13.78 & -4.38 & -4.38 & 0.048 & 5 & 0.980 & 536.25 & -4.26 $\pm$0.87 \\ 
        HD 165341 &  K0V & 4.03 & 0.86 & 5407 & 0.85 & 5.10 & -4.66 & -4.65 & 0.032 & 4 & 6.783 & 1099.00 & -4.79 $\pm$0.74 \\ 
        HD 189733 &  K0V+M4V & 7.65 & 0.93 & 5015 & 0.72 & 19.78 & -4.50 & -4.50 & 0.007 & 3 & 0.242 & 72.33 & -4.52 $\pm$0.89 \\ 
        HD 128621 &  K1V & 1.33 & 0.88 & 5336 & 0.74 & 1.25 & -5.02 & -5.02 & 0.043 & 3 & 81.549 & 5088.15 & -5.20 $\pm$0.94 \\ 
        HD 114783 &  K1V & 7.56 & 0.93 & 5100 & 0.77 & 21.08 & -5.01 & -5.01 & 0.064 & 5 & 0.263 & 11.22 & -5.37 $\pm$0.96 \\ 
        HD 97658 &  K1V & 7.71 & 0.85 & 5192 & 0.71 & 21.58 & -4.99 & -4.97 & 0.059 & 5 & 0.228 & 4.61 & -5.69 $\pm$0.98 \\ 
        HD 103095 &  K1V & 6.45 & 0.75 & 5265 & 0.53 & 9.18 & -4.87 & -4.85 & 0.029 & 5 & 0.730 & 1.25 & -6.77 $\pm$1.15 \\ 
        HR 1925 &  K1V & 6.23 & 0.84 & 5303 & 0.77 & 12.28 & -4.53 & -4.55 & 0.042 & 5 & 0.894 & 261.89 & -4.53 $\pm$0.89 \\ 
        HD 40307 &  K2.5V & 7.15 & 0.95 & 4893 & 0.62 & 12.94 & -4.84 & -4.84 & 0.014 & 2 & 0.384 & 4.61 & -5.92 $\pm$1.00 \\ 
        HD 192263 &  K2.5V & 7.77 & 0.96 & 4996 & 0.68 & 19.65 & -4.55 & -4.55 & 0.019 & 4 & 0.217 & 58.61 & -4.57 $\pm$0.90 \\ 
        epsilon Eri &  K2V & 3.73 & 0.88 & 4975 & 0.72 & 3.20 & -4.46 & -4.47 & 0.040 & 10 & 8.942 & 1884.00 & -4.68 $\pm$0.63 \\ 
        HD 192310 &  K2V & 5.72 & 0.91 & 5091 & 0.76 & 8.80 & -5.04 & -5.05 & 0.029 & 3 & 1.426 & 61.11 & -5.37 $\pm$0.96 \\ 
        HD 99492 &  K2V & 7.53 & 1.02 & 4801 & 0.77 & 18.21 & -4.87 & -4.88 & 0.024 & 6 & 0.270 & 22.45 & -5.08 $\pm$0.94 \\ 
        HD 22468 &  K2V & 5.71 & 0.92 & 4833 & 2.84 & 29.60 & -3.84 & -3.84 & 0.038 & 2 & 1.443 & 7482.57 & -3.29 $\pm$0.81 \\ 
        HD 155886 &  K2V & 5.08 & 0.85 & 5089 & 0.69 & 5.96 & -4.58 & -4.59 & 0.017 & 5 & 2.579 & 236.95 & -5.04 $\pm$0.93 \\ 
        HD 128311 &  K3V & 7.45 & 1.00 & 4863 & 0.69 & 16.34 & -4.45 & -4.45 & 0.048 & 4 & 0.292 & 106.00 & -4.44 $\pm$0.89 \\ 
        HD 104067 &  K3V & 7.92 & 0.98 & 4974 & 0.67 & 20.38 & -4.75 & -4.76 & 0.081 & 6 & 0.188 & 37.41 & -4.70 $\pm$0.91 \\ 
        HD 156668 &  K3V & 8.42 & 1.01 & 4832 & 0.67 & 24.35 & -5.01 & -5.01 & 0.038 & 1 & 0.119 & 4.99 & -5.38 $\pm$0.96 \\ 
        WASP 69 &  K5V & 9.87 & 1.06 & 4875 & 0.69 & 50.03 & -4.54 & -4.54 & 0.038 & 1 & 0.031 & 7.11 & -4.64 $\pm$0.90 \\ 
        HD 85512 &  K6V & 7.65 & 1.18 & 4421 & 0.53 & 11.28 & -4.92 & -4.93 & 0.097 & 6 & 0.242 & 8.60 & -5.45 $\pm$0.96 \\ 
        GJ 832 &  M1.5V & 8.67 & 1.50 & 3993 & 0.18 & 4.96 & -5.07 & -5.07 & 0.038 & 1 & 0.094 & 59.11 & -4.20 $\pm$0.88 \\ 
        GJ 667 C &  M1.5V & 10.22 & 1.57 & 3775 & 0.14 & 7.25 & -4.66 & -4.67 & 0.049 & 5 & 0.023 & 12.12 & -4.27 $\pm$0.88 \\ 
        LTT 2050 &  M1V & 10.33 & 1.51 & 3914 & 0.20 & 11.21 & -4.99 & -4.99 & 0.093 & 2 & 0.020 & 28.63 & -3.85 $\pm$0.85 \\ 
        HD 197481 &  M1V & 8.63 & 1.42 & 3992 & 0.36 & 9.72 & -3.91 & -3.88 & 0.079 & 3 & 0.098 & 1526.40 & -2.81 $\pm$0.61 \\ 
        GJ 176 &  M2.5V & 9.95 & 1.54 & 3898 & 0.20 & 9.47 & -4.75 & -4.75 & 0.043 & 4 & 0.029 & 52.20 & -3.75 $\pm$0.85 \\ 
        GJ 3470 &  M2V & 12.33 & 1.17 & 3776 & 0.22 & 29.45 & -4.86 & -4.86 & 0.038 & 1 & 0.003 & 50.52 & -2.81 $\pm$0.79 \\ 
        GJ 436 &  M3.5V & 10.61 & 1.45 & 3660 & 0.17 & 9.76 & -5.09 & -5.09 & 0.001 & 2 & 0.016 & 16.17 & -3.99 $\pm$0.86 \\ 
        AD Leo &  M4V & 9.52 & 1.30 & 3859 & 0.13 & 4.97 & -3.99 & -3.99 & 0.007 & 2 & 0.043 & 2332.00 & -2.27 $\pm$0.74 \\ 
        EV Lac &  M4V & 10.26 & 1.59 & 3742 & 0.10 & 5.05 & -3.75 & -3.75 & 0.038 & 1 & 0.022 & 954.00 & -2.36 $\pm$0.55 \\ 
        Proxima Centauri &  M5.5V & 11.13 & 1.82 & 3296 & 0.02 & 1.30 & -4.29 & -4.29 & 0.012 & 2 & 0.010 & 318.00 & -2.49 $\pm$1.52 \\ 
        GJ 876 &  M5V & 10.19 & 1.56 & 3837 & 0.09 & 4.68 & -5.00 & -5.00 & 0.000 & 2 & 0.023 & 180.19 & -3.11 $\pm$0.80 \\ 
       \end{longtable}
\end{landscape}
}

\end{document}